\documentclass[%
 reprint,
superscriptaddress,
showpacs,
 amsmath,amssymb,
 amsfonts, 
 aps,
prb,
floatfix,
]{revtex4-1}

\pdfoutput=1

\usepackage{float}
 
\usepackage{graphicx}% Include figure files
\usepackage{dcolumn}% Align table columns on decimal point
\usepackage{bm}% bold math
\usepackage{array}
\usepackage{makecell}
\usepackage{multirow}
\usepackage{adjustbox}
\newcolumntype{K}[1]{>{\centering\arraybackslash}p{#1}}
\newcolumntype{D}[1]{>{\arraybackslash}p{#1}}

\usepackage{xcolor}

\newcommand{\red}{\textcolor{black}}    % Uncomment to restore black color

\usepackage[normalem]{ulem}
\usepackage{cancel}

\begin{document}

\title{Space-charge effects in the field-assisted thermionic emission from nonuniform cathodes}

\author{Anna Sitek}
 \email{anna.sitek@pwr.edu.pl}
 \affiliation{Department of Engineering, Reykjavik University, 
	          Menntavegur 1, IS-102 Reykjavik, Iceland}
 \affiliation{Department of Theoretical Physics, Wroclaw University of Science and Technology, 
              Wybrze\.{z}e  Wyspia\'{n}skiego  27, 50-370 Wroclaw, Poland}

\author{Kristinn Torfason}
 \affiliation{Department of Engineering, Reykjavik University, 
  	          Menntavegur 1, IS-102 Reykjavik, Iceland} 
 
\author{Andrei Manolescu}
\affiliation{Department of Engineering, Reykjavik University, 
	         Menntavegur 1, IS-102 Reykjavik, Iceland}

\author{\'Ag\'ust Valfells}
 \affiliation{Department of Engineering, Reykjavik University, 
              Menntavegur 1, IS-102 Reykjavik, Iceland}

\begin{abstract} 
We use computational simulations to study the electron emission and propagation in planar vacuum diodes.
We show how space-charge affects thermionic emission from cathodes with two different values of work function that form a checkerboard pattern of finite extent on the cathode surface. We confirm that, for intermediate cathode temperature, the local current density from low work function regions can exceed the space-charge limit for the entire cathode. As the cathode temperature rises space-charge effects lead to homogeneous current density from the interior of the emitting area and a higher current density from its periphery. We show how beam emittance and brightness are affected, and show that the operating temperature for optimal brightness is such that it corresponds to the transition region between source-limited and space-charge limited emission. Finally we show how beam current and beam quality depend on how fine grained the structure of the cathode is.   
\end{abstract}

%\pacs{73.21.La, 73.22.Dj, 78.67.Hc}

\maketitle

\section{\label{sec:introduction} Introduction}

Cathodes, used as electron sources in a large number of devices \cite{Dowell10,Zhang17a}, are generally inhomogeneous at the microscale whether by design or happenstance. In particular, the work function may vary along the cathode. This may be due to, for example, the cathode being made of many crystal grains with different facets facing the vacuum gap, the presence of adsorbates on the surface, or because of inherent structure such as in the case of LaB\(_6\) cathodes.  Since the work function is an important determinant in the emission rate for thermionic emission, field emission, and photoelectric emission, it follows that any variation in the work function on the cathode is likely to lead to variations in the current density over the cathode. This variation may, in turn, have an effect on the operational characteristics of diodes and the quality of electron beams drawn from the cathode. 

Thermionic cathodes are commonly used in electron devices as they are relatively inexpensive and robust. The current in a thermionic diode is temperature dependent, and in the source-limited regime it is simply described by the Richardson-Laue-Dushman (RLD) \cite{Richardson16,Dushman23} law that relates the current density to the work function and cathode temperature. As the current density increases with temperature, space charge decreases the electric field at the cathode surface and the diode enters the space-charge-limited regime in which the current density is traditionally described by the Child-Langmuir law that depends on the applied voltage and spacing of the diode. The characteristic curve relating the diode current to the cathode temperature is called a Miram curve and shows a smooth transition (a "knee" in the curve) from the source-limited current to the space-charge-limited current that cannot be explained by the simple forms of the Richardson-Dushman and Child-Langmuir laws. To ensure a long cathode lifetime, efficient operation and acceptable beam characteristics, it is necessary for the designer or operator of a  device, based on thermionic emission, to have a firm understanding of the Miram curve for the cathode.
Recent work on nonuniform thermionic emission has offered some clarity on the issue. In particular, a paper by Chernin \textit{et al.} \cite{Chernin20} has advanced understanding of how the Miram curve is affected via mutual space-charge effects due to current emanating from different regions of the cathode, and offers a useful method of calculating the current from a thermionic cathode with periodically varying work function that compares favorably with particle-in-cell simulations.

In this paper we use molecular dynamics-based simulations to further investigate thermionic emission from cathodes with a nonuniform work function. As well as investigating the effects of inhomogeneity on the Miram curve, we examine how it affects the beam quality, i.e., emittance and brightness. Although this work uses a combined thermionic-field emission model and a different characteristic length scale than the work of Chernin \textit{et al.} \cite{Chernin20}, the core physics is similar enough to offer a comparison with their work, showing commonalities, some differences, and further insight into how microstructure of the cathode is manifested in the beam characteristics. The physics of thermionic emission from an inhomogeneous cathode will also be examined in the context of some recent work on space-charge-influenced emission from discrete and irregular emitters.   

%----------------------------------
The paper is organized as follows. In the next section we describe the system under study and the computational method. In Sec.\ \ref{sec:results} we present our results. In particular, in Sec.\  \ref{sec:uni_non_uni} we compare the currents obtained from uniform and nonuniform cathodes. In the following sections we focus on the nonuniform cathodes with a checkerboard work function pattern. 
\red{We} show how the current, emittance and brightness change with increasing temperature (Sec.\ \ref{sec:checks}) and grid fineness (Sec.\ \ref{sec:grid}). Finally, in Sec.\ \ref{sec:conclusions} we summarize the results.

\section{\label{sec:model} Model}

\begin{figure}[tb]
	\centering
	\includegraphics[scale=0.67]{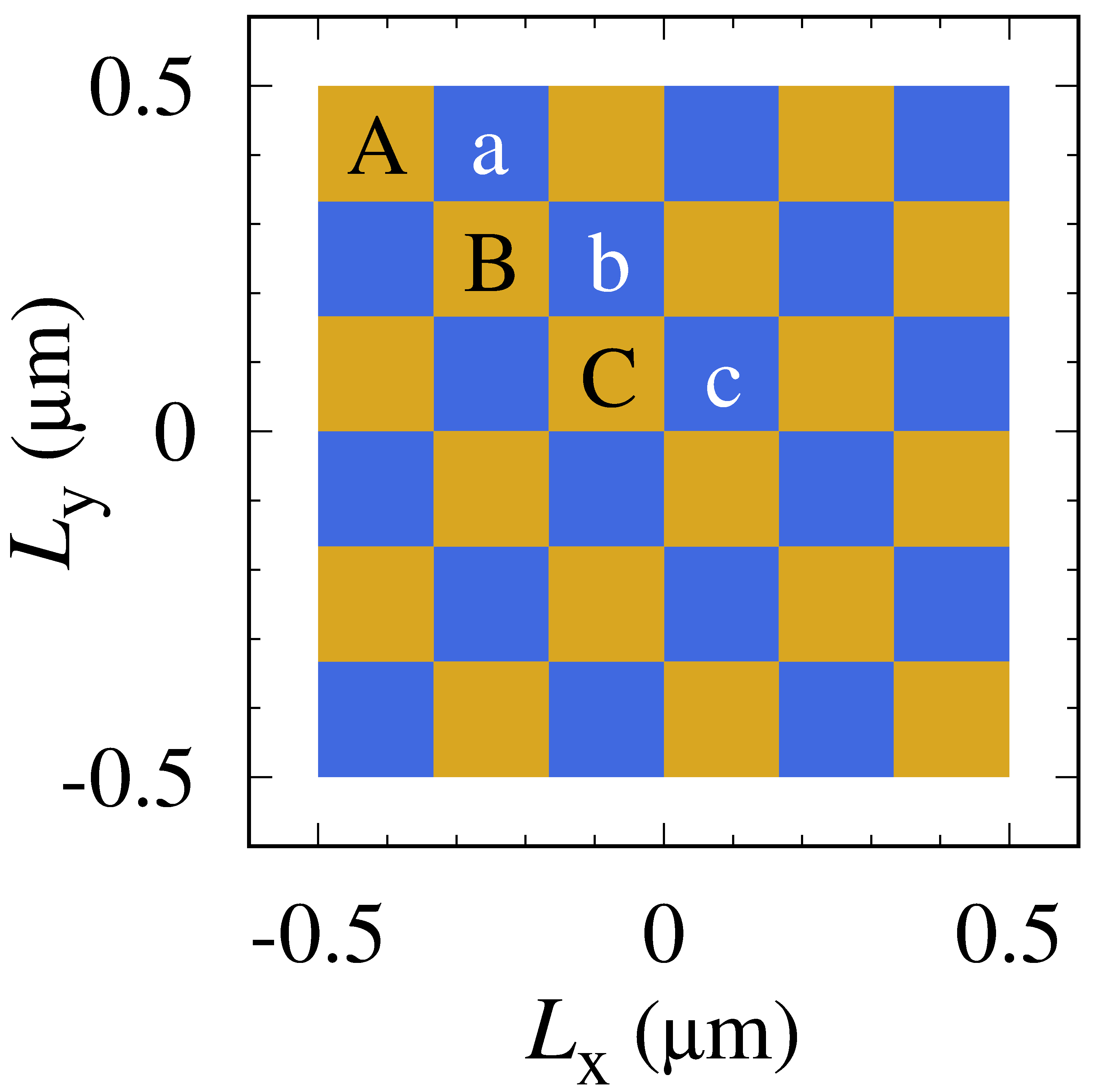}
	\caption{Work function on the \red{emitting area of the} cathode surface. Yellow indicates  areas of lower 
		work function $\Phi_1=2.0$ eV, while blue represents regions where the
		work function takes the higher value $\Phi_2=2.5$ eV.
		The letters indicate the checks for which the partial currents are shown in 
		Fig.\ \ref{Fig_Diagonal}.}
	\label{Fig_Cathode}
\end{figure}

The system under study is a planar vacuum diode of infinite extent. The cathode is nonemitting, except for thermionic emission from a square patch of side length $L$. 
The emitting area on the cathode is characterized by a work function that has a checkerboard pattern built of alternating high \red{and} low work function values, as shown in  Fig.\ \ref{Fig_Cathode}. The number of checks can be varied to form an $N$-by-$N$ pattern. 
\red{We do not apply any constraints on the size of the anode, and thus it can absorb every electron reaching its surface.}

We study the field-assisted thermal emission using the method developed by 
Jensen \cite{Kevin}. The current density injected into the system is described with the 
general formula
\begin{eqnarray}
\label{func_J}	
J(F,T) = A_{\mathrm{RLD}}T^2 n \int\limits^{\infty}_{-\infty}
\frac{\ln\left[1+e^{n\left(x-s\right)}\right]}{1+e^x}dx, 
\end{eqnarray}   
where $F$ is  \red{the electrostatic force due to} the field \red{acting on an electron} at a \red{point on the} cathode surface. It depends on the bias voltage applied between the electrodes ($V_0$) and the space charge due to the electrons in the gap.
Here $T$ stands for temperature,
\begin{eqnarray*}
	A_{\mathrm{RLD}} = \frac{qmk_{\mathrm{B}}^2}{2\pi^2\hbar^3}
\end{eqnarray*} 
is the Richardson constant, $q$ and $m$ are the electron charge and mass, $k_{\mathrm{B}}$ and $\hbar$ represent the Boltzmann and Planck constants, respectively, and
\begin{eqnarray*}
	n(F,T) = \frac{\beta_T}{\beta_F}
\end{eqnarray*} 
is the ratio between the temperature energy slope factor $\beta_T=1/{k_{\mathrm{B}}T}$
and its field counterpart, \red{i.e., the field energy slope factor} $\beta_F$. 
\red{For the thermal dominated regime, it takes the form \cite{Kevin}}
\begin{eqnarray*}
	\red{\beta_F = \frac{\pi}{\hbar F} \sqrt{m\Phi\sqrt{\mathit{l}}}},
\end{eqnarray*}	
where \red{$\Phi$ stands for the work function} and
\begin{eqnarray*}
	\mathit{l}=\frac{q^2}{4\pi\epsilon_0}\frac{F}{\Phi^2} 
\end{eqnarray*}	
\red{is a dimensionless parameter.}
Finally, 
\red{the $s$ function in Eq.\ \ref{func_J} is}
\begin{eqnarray*}
	\red{s(F,T) = \beta_F \phi},
\end{eqnarray*} 
\red{with $\phi = (1-\sqrt{l})\Phi$ the image charge reduced barrier}.

%%%%%%%%%%%%%%%%%%%%%%%%%%%%%%%%%%%%%%%%%%%%%%%%%%%%%%%%%%%%%%%%%%%%%%%%%%%
%%%%%%%%%%%%%%%%%%%%%%%%%%%%%%%%%%%%%%%%%%%%%%%%%%%%%%%%%%%%%%%%%%%%%%%%%%%

Equation (\ref{func_J}) describes all three emission regimes, i.e., field and thermal emissions as well as the intermediate range where the impact of both processes is comparable. \red{However, the forms of $\beta_F$ and the function $s(F,T)$ depend on the regime chosen.} In this paper we \red{focus on} the field-assisted thermal emission; in this regime the particles are extracted from the cathode due to the thermal energy, while the external electric field serves the twofold purpose of reducing the surface barrier via the Schottky effect and sweeping electrons away from the cathode.

After determining the current density injected into the system (Eq.\ \ref{func_J}) we use the Metropolis-Hastings algorithm to find favorable spots for the electrons to be released from the cathode surface. Further, we 
perform high-resolution molecular dynamics simulations of particle advancement that include full Coulomb interactions of electrons and we determine the current and beam quality features, such as emittance and brightness.

The current is calculated according to the Ramo-Shockley theorem \cite{Ramo39},
\begin{equation*}
\red{I = \frac{q}{d}\sum_i \left(v_{z}\right)_i,} 
\end{equation*}
where \red{$d$ is the gap spacing and} $v_z$ the $z$ component of the electron's instantaneous velocity.  The summation is carried out over the contributions from all electrons in the gap.

The lateral spread of the electron beam in the $x$ direction and its disorder are characterized by the statistical emittance \cite{Reiser94, buon1992beam},
\begin{eqnarray*}
	\epsilon_x = \sqrt{\left< x^2 \right> \left< x^{\prime 2} \right> - \left< xx^\prime \right>^2}\ ,
\end{eqnarray*}
where $x$ denotes the position of an electron along the $x$ axis and $x'$ is the slope of the trajectory. The emittance for the $y$ axis, $\epsilon_y$, is calculated analogously. In our model the $z$ axis originates at
\red{the cathode and is normal to its surface.}

Another important parameter describing the beam quality is the brightness, which measures current density per unit solid angle. As a measure of brightness, $B$, we use \cite{reiser1994theory}
\begin{eqnarray*}
	B = \frac{2I}{\pi^2\epsilon_x\epsilon_y}.
\end{eqnarray*}
%
%

%%%%%%%%%%%%%%%%%%%%%%%%%%%%%%%%%%%%%%%%%%%%%%%%%%%%%%%%%%%%%%%%%%%%%%%%%%%
\begin{figure}[tb]
	\includegraphics[scale=0.67]{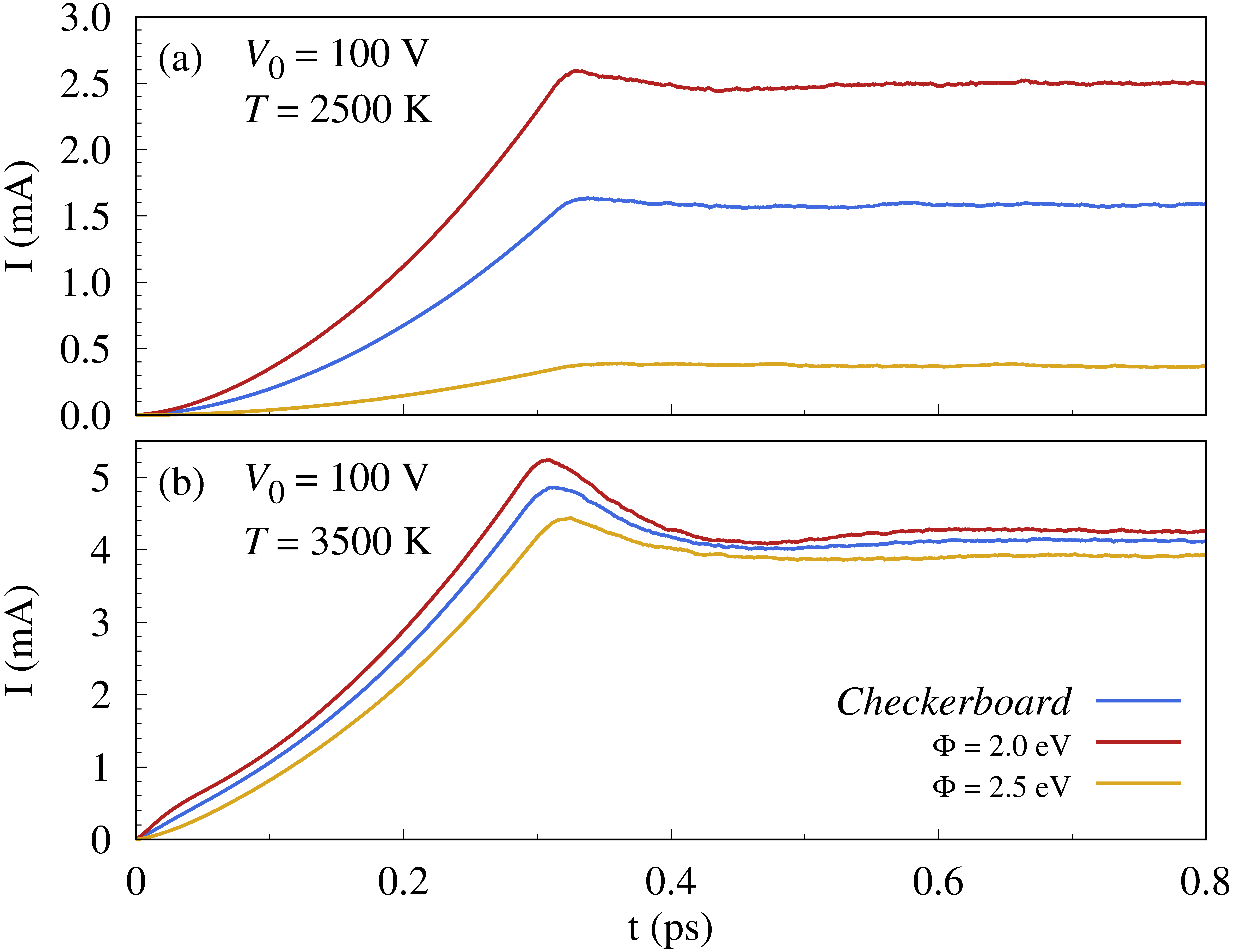}
	\caption{Evolution of the current for cathodes with uniform 
		work functions ($\Phi$; red and yellow lines) and for a checkerboard 
		cathode with $N=6$ (blue line). The applied voltage is set to $V_0=100$ V and 
		the temperatures are (a) $T=2500$ K and (b) $T=3500$ K. The line 
		description shown in panel (b) is valid for both panels.}
	\label{Fig_time}
\end{figure}
%%%%%%%%%%%%%%%%%%%%%%%%%%%%%%%%%%%%%%%%%%%%%%%%%%%%%%%%%%%%%%%%%%%%%%%%%%%

%%%%%%%%%%%%%%%%%%%%%%%%%%%%%%%%%%%%%%%%%%%%%%%%%%%%%%%%%%%%%%%%%%%%%%%%%%%
\begin{figure}[tb]
	\includegraphics[scale=0.67]{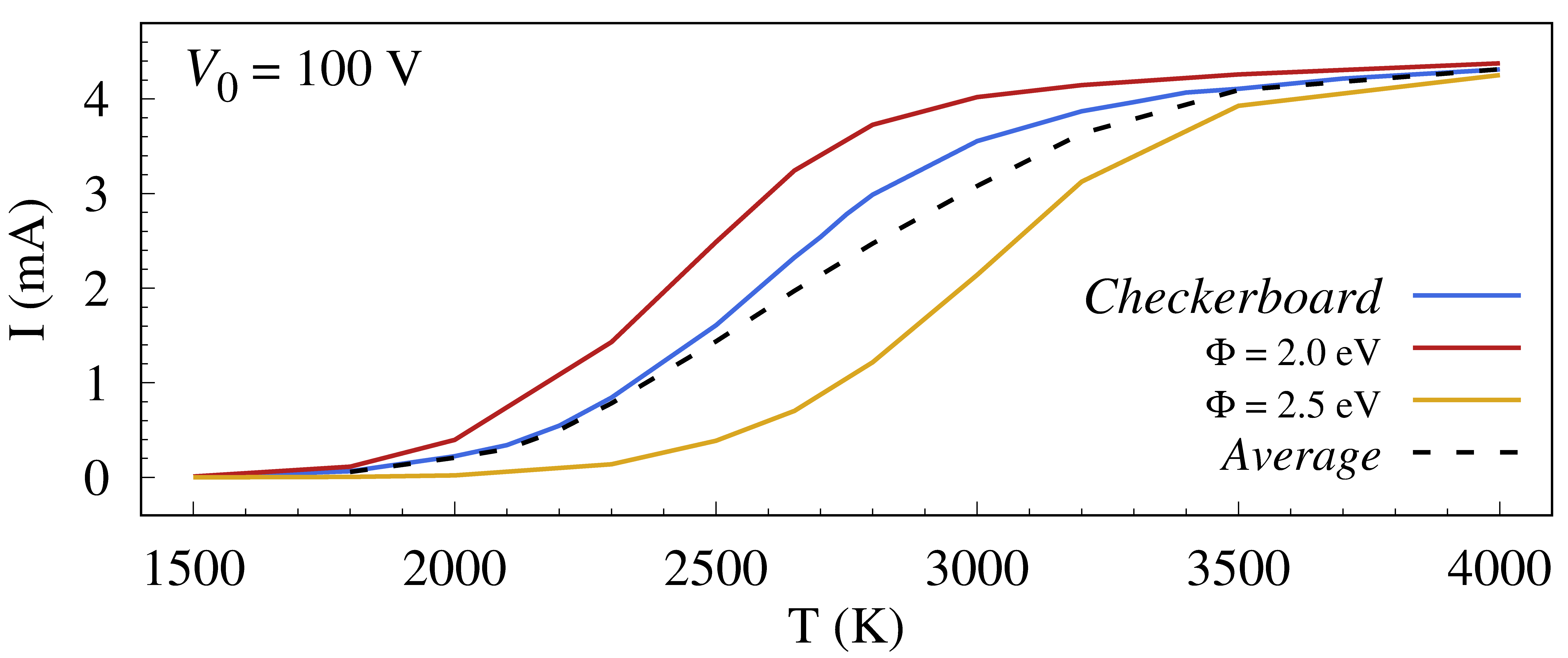}
	\caption{Steady-state current versus temperature for cathodes with uniform 
		work functions ($\Phi$; red and yellow lines) and for a checkerboard 
		cathode with $N=6$ (blue line). The black dahed line represents the average of currents 
		from the uniform cathodes. The applied voltage is set to $V_0=100$ V.}
	\label{Fig_Full_Ch}
\end{figure}
%%%%%%%%%%%%%%%%%%%%%%%%%%%%%%%%%%%%%%%%%%%%%%%%%%%%%%%%%%%%%%%%%%%%%%%%%%%

%%%%%%%%%%%%%%%%%%%%%%%%%%%%%%%%%%%%%%%%%%%%%%%%%%%%%%%%%%%%%%%%%%%%%%%%%%%
\begin{figure*}[tb]
	\centering
	\includegraphics[width=\textwidth]{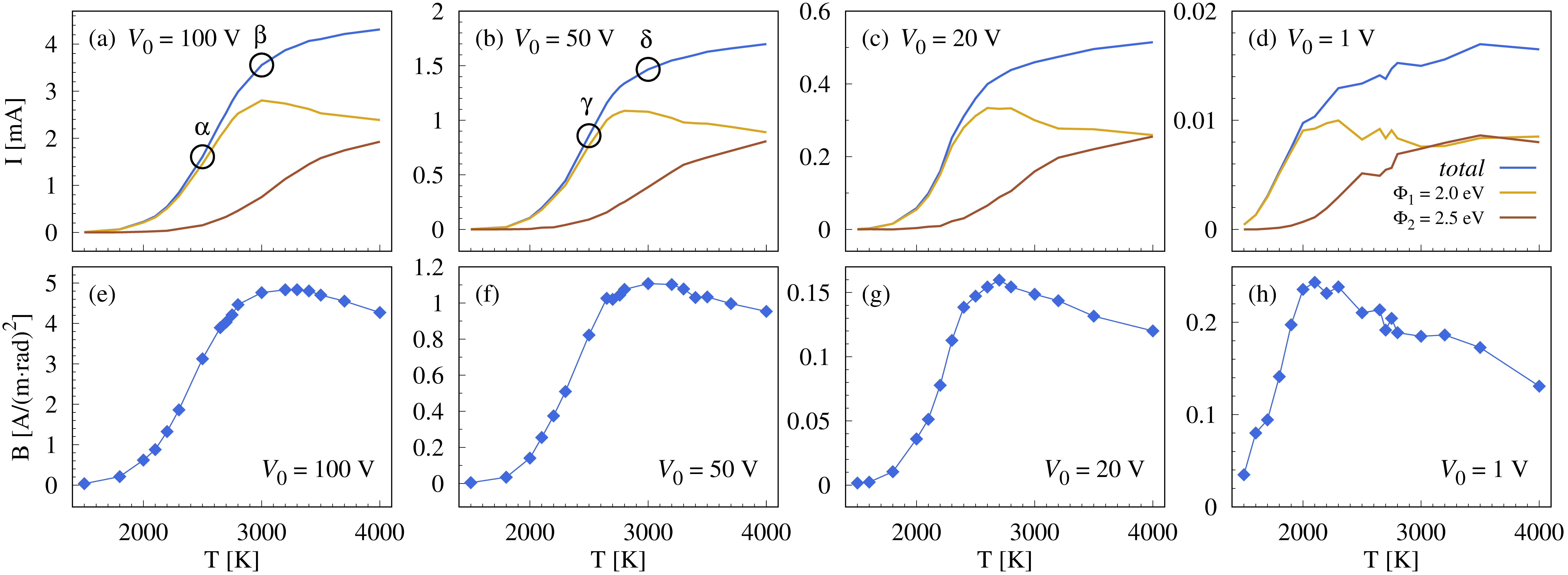}
	\caption{Total steady-state current and its contributions from areas of lower ($\Phi_1=2.0$ eV) and higher ($\Phi_2=2.5$ eV) work functions versus temperature for a checkerboard cathode with $N=6$ [(a)-(d)] and the corresponding brightness [(e)-(h)]. 
		The applied bias voltages are (a),(e) $V_0=100$~V, (b),(f) $V_0=50$~V, (c),(g)
		$V_0=20$~V, and (d),(h) $V_0=1$~V. The line description shown in panel (d) is valid for all panels in the upper row. \red{The Greek letters in (a) and (b) show points on the Miram curves corresponding to the values of temperature and applied voltage for which the dependence on the grid fineness of the work function is studied in Sec.\ \ref{sec:grid}.}}
	\label{Fig_I_B}
\end{figure*}
%%%%%%%%%%%%%%%%%%%%%%%%%%%%%%%%%%%%%%%%%%%%%%%%%%%%%%%%%%%%%%%%%%%%%%%%%%%

In this paper we show some pertinent results from simulations performed for emitters of side length $L=1$ $\mu$m 
in a vacuum diode of gap spacing $d=1$ $\mu$m. We calculate the currents from emitters of uniform work function, but we focus on structures with 
two values of work function on the emitting region of the cathode, which we set to $\Phi_1=2.0$ eV and $\Phi_2=2.5$ eV.
The corresponding areas form a checkerboardlike pattern consisting
of $N\in (2; 48)$ alternating squares in each row and column, see Fig.\ \ref{Fig_Cathode}.

It is important to note that in the following simulations, the cathode temperature may be driven as high as 4000 K, which is clearly beyond any realistic temperature range for actual cathodes. The reason for doing this is that this work is part of a series of investigations on the effects of charge discreteness, cathode inhomogeneity, and small emitter size being undertaken by our group. In that work the diode system we have been investigating is of emitter side length and diode spacing of the order of 1 $\mu$m and with the work function values described above \cite{Torfason15,Torfason20}. \red{The reader should also note that, even though the emitting area is of small dimension, it is planar and embedded in an infinite planar cathode. Thus, there is no enhancement of the applied electric field by the emitter.}
In order to get good statistics and drive the current well into the space-charge limit, while using the aforementioned parameters, it was necessary to allow the cathode temperature to become as high as 4000 Kelvin.
As we will later demonstrate, this does not alter the main findings of this paper which can be understood to apply to more realistic thermionic cathodes as well.

\section{\label{sec:results} Results}

In this section we present the results of our simulations of field-assisted thermal emission
from cathodes with uniform and nonuniform work functions. In particular, we compare Miram curves 
obtained from both types of structures, and we show how the tiling of the emitting area affects the transverse structure and quality of the beam emanating from the cathode as a function of applied potential and cathode temperature.

\subsection{\label{sec:uni_non_uni} Comparison of uniform and checkerboard cathodes}

Thermionic emission from a cathode is strongly influenced by the work function. In Figs.\ \ref{Fig_time} and\ \ref{Fig_Full_Ch} we compare the thermionic current in a diode for three different work function configurations: a uniform work function of 2.0 eV, a uniform work function of 2.5 eV, and a $6$-by-$6$ checkerboard structure where the work function alternates between the two aforementioned values. All other determining parameters are held the same. 
\red{For $t<0$, the bias of the applied voltage is such that it suppresses any current from the cathode. At $t=0$ the applied potential is reversed and set to a finite value that allows the electrons to travel from the cathode to the anode.}
In both graphs shown in Fig.\ \ref{Fig_time} we see how the current increases with time as the initially emitted electrons traverse the diode gap. Once the gap has been filled, the current settles to a steady state. For a temperature of 2500 K, as shown in the upper graph of Fig.\ \ref{Fig_time}, the steady-state current is source limited and the current for the 2.5 eV work function cathode is nearly an order of magnitude smaller than that for the 2.0 eV cathode. At 3500 K the steady-state current is independent of the work function, indicating that it is now space-charge limited. Note that the current can momentarily exceed the space-charge limit as the initial burst of electrons reaches the anode after approximately $0.3$ ps.

The temperature dependence of the steady-state current for the three systems is shown in
Fig.\ \ref{Fig_Full_Ch}. The three Miram curves show the characteristic transition from the Richardson-Laue-Dushman current to the space-charge-limited current. Interestingly, the characteristic "knee", in the transition region between the two current regimes, is present even when the work function is uniform. This is different from the results reported by Chernin \textit{et al.} \cite{Chernin20} for a uniform work function where a sharp transition takes place, and the "knee" appears only when the cathode has a nonuniform work function. However, their results were based on an infinite emitting area whereas our simulations are based on a finite emitter. This indicates that the Miram curve may also be affected by edge effects, as is known to be the case for space-charge-limited emission and field emission from finite area emitters. We also note that the \red{emitting} surface of the checkerboard cathode is equally divided between areas associated with each value of the work function. Nonetheless, the total current can be significantly larger than the algebraic average of the currents from the uniform systems, at least for $N=6$.

\subsection{\label{sec:checks} Temperature effects on a checkerboard surface}

%%%%%%%%%%%%%%%%%%%%%%%%%%%%%%%%%%%%%%%%%%%%%%%%%%%%%%%%%%%%%%%%%%%%%%%%%%%
\begin{figure}[tb]
	\centering
	\includegraphics[scale=0.67]{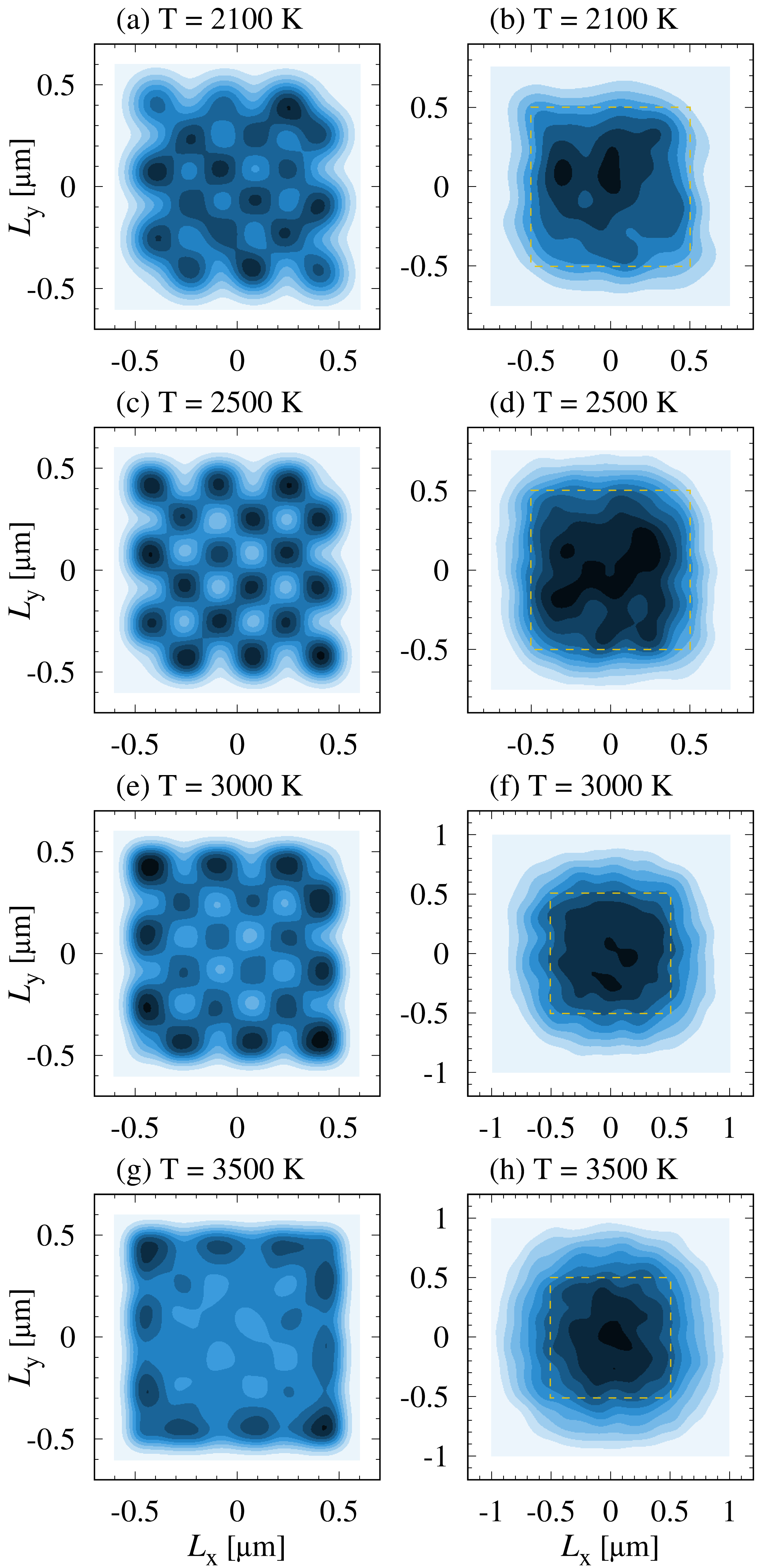}
	\caption{Local density of electrons at the cathode (left column) and anode (right column)
		for $V_0=50$ V and for temperatures (a),(b) $T=2100$ K, (c),(d) $T=2500$ K, (e),(f) $T=3000$ K, and (g),(h) $T=3500$ K. The yellow dashed lines in the right column indicate the size and position of the cathode.}
	\label{Fig_C_A}
\end{figure}
%%%%%%%%%%%%%%%%%%%%%%%%%%%%%%%%%%%%%%%%%%%%%%%%%%%%%%%%%%%%%%%%%%%%%%%%%%%

%%%%%%%%%%%%%%%%%%%%%%%%%%%%%%%%%%%%%%%%%%%%%%%%%%%%%%%%%%%%%%%%%%%%%%%%%%%
\begin{figure}[tb]
	\includegraphics[scale=0.67]{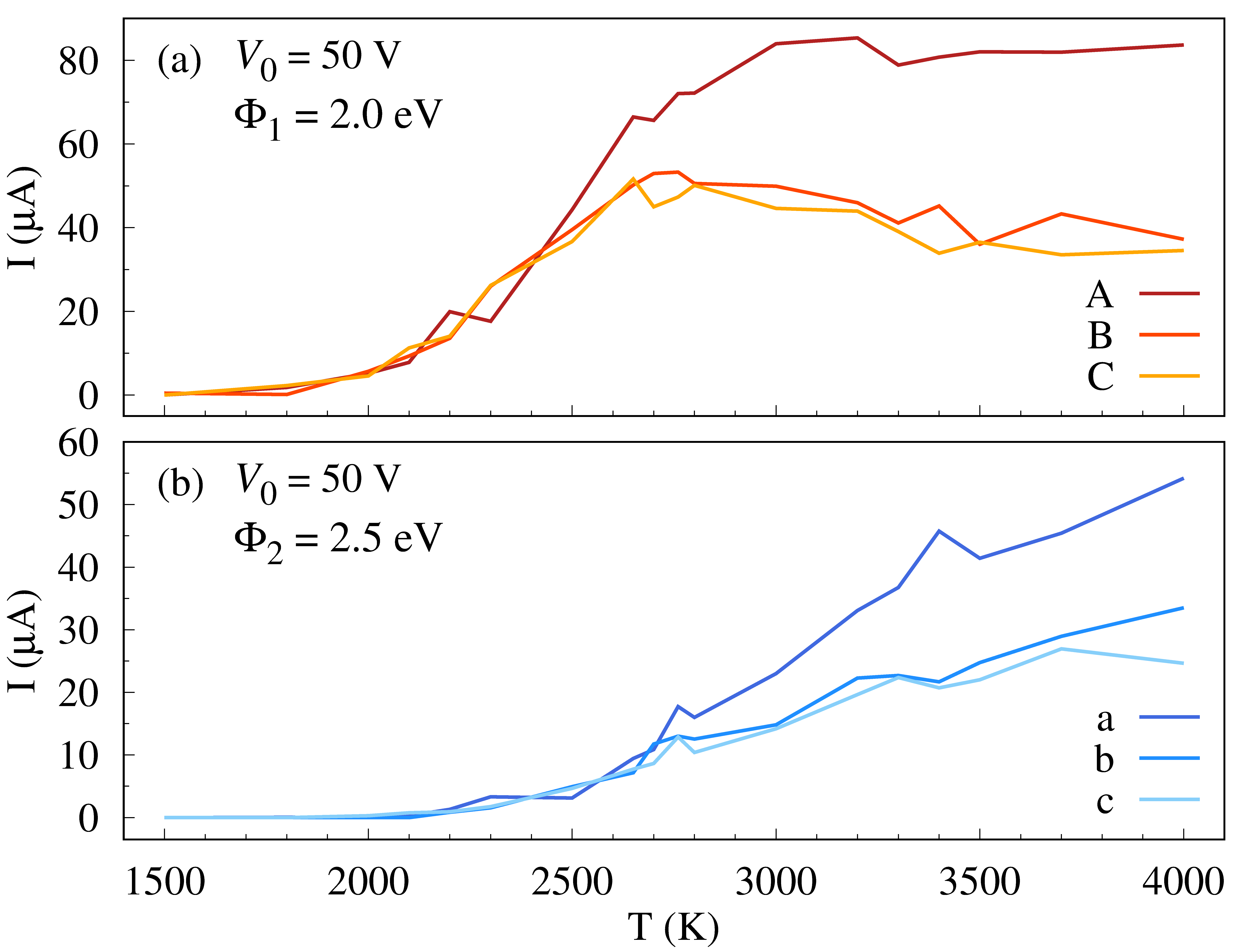}
	\caption{Steady-state currents emitted from single areas of lower work function ($\Phi_1=2.0$ eV) situated along 
		the diagonal (a) and originating from the neighboring areas of higher 
		work function ($\Phi_2=2.5$ eV) (b) for the applied voltage $V_0=50$ V. 
		The positions of the particular checks (A, B, C, a, b, c) 
		are defined in Fig.\ \ref{Fig_Cathode}.}
	\label{Fig_Diagonal}
\end{figure}
%%%%%%%%%%%%%%%%%%%%%%%%%%%%%%%%%%%%%%%%%%%%%%%%%%%%%%%%%%%%%%%%%%%%%%%%%%%

%%%%%%%%%%%%%%%%%%%%%%%%%%%%%%%%%%%%%%%%%%%%%%%%%%%%%%%%%%%%%%%%%%%%%%%%%%%
\begin{figure}[tb]
	\includegraphics[scale=0.67]{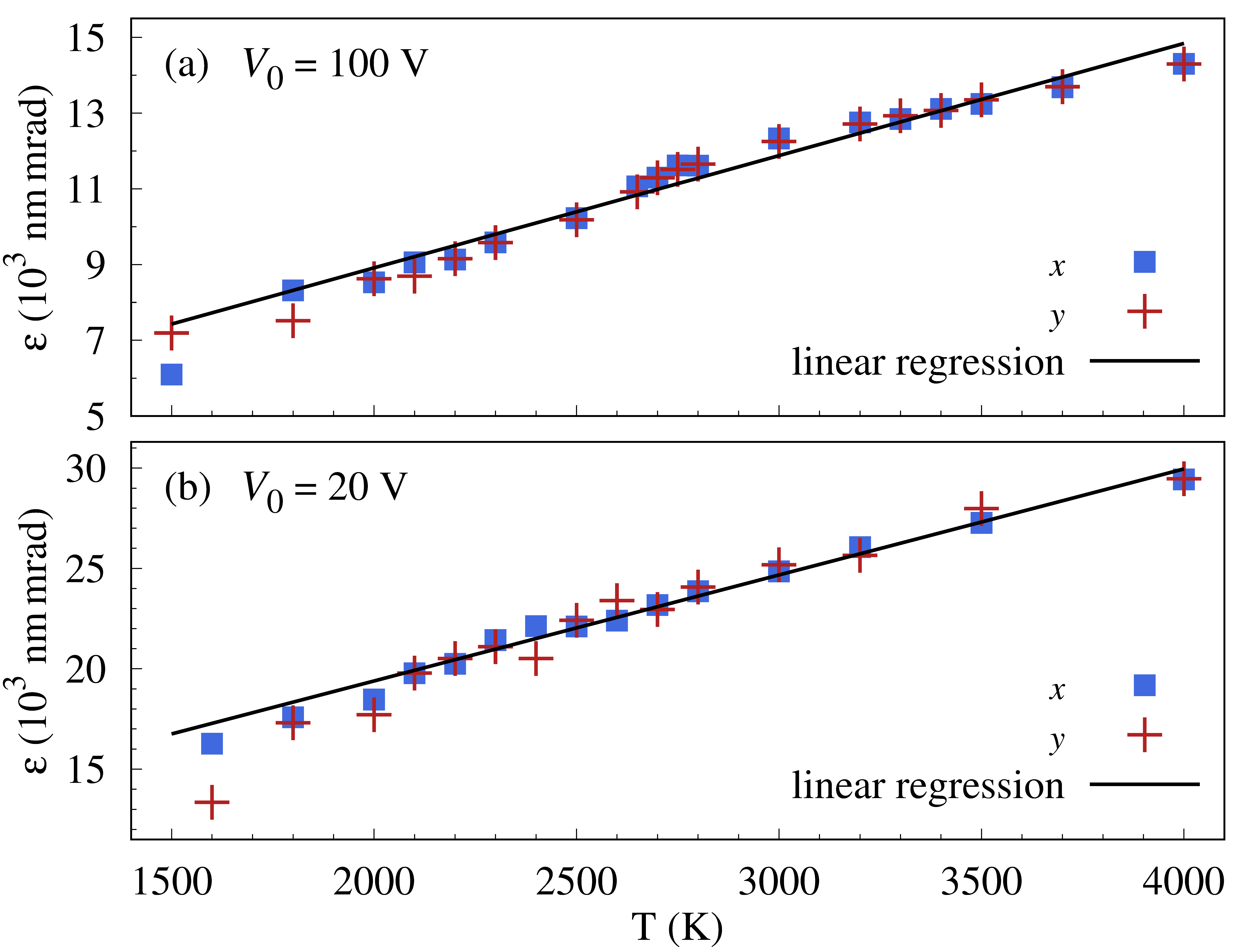}
	\caption{Emittance ($x$ and $y$ coordinates) versus temperature for the checkerboard with $N=6$ and applied voltages (a) $V_0=100$ V and (b) $V_0=20$ V. The linear regression is calculated for $T>2000$ K.}
	\label{Fig_E}
\end{figure}
%%%%%%%%%%%%%%%%%%%%%%%%%%%%%%%%%%%%%%%%%%%%%%%%%%%%%%%%%%%%%%%%%%%%%%%%%%%

The coexistence of checks with two distinct work functions results in variation in the current density across the emitter. This can be seen in the  
left column of Fig.\ \ref{Fig_C_A}, which shows the electron density at the cathode surface for different values of the cathode temperature. For lower temperatures the electron density reflects the underlying structure of the work function shown in Fig.\ \ref{Fig_Cathode}.
Consequently, the currents originating from the areas of different work function contribute differently to the total current.  
In Figs.\ \ref{Fig_I_B}(a)-\ref{Fig_I_B}(d) 
we look separately at the current coming from areas of higher and lower work functions, respectively. For the applied 
voltages used the current turns on only when the cathode is heated to 1500 K. For lower temperature, 
electrons are emitted solely from the areas of lower work function. With increasing temperature, the areas of higher 
work function start contributing to the emission. As current rises with increasing temperature and the space-charge limit is reached, the current is evenly distributed between the high and low work function areas. Interestingly, the current from the low work function region reaches a maximum value before decreasing as more of the current comes from the high work function regions. This is in accordance with results obtained by Chernin \textit{et al.} \cite{Chernin20} and is readily explained as follows. When the temperature is so low that the current density from high work function checks is negligible, the low work function areas do not experience space-charge effects from their nearest neighbors. Thus, the current density from them may be higher than that predicted by the classic Child-Langmuir law, as has been shown in previous studies \cite{Lau01,Luginsland96}. However, as the temperature increases, the current density from the high work function areas increases as well. Thus, the space-charge effect from that current on the neighboring low work function checks increases in turn. Consequently, the current density from the low work function areas is reduced. Further increasing the temperature enhances this space-charge interaction until the current density across the interior of the emitting surface becomes uniform. Note that, as the gap voltage is decreased (and subsequently the space-charge-limited current) the temperature, at which the current from the low work function region peaks, is reduced. The temperature at which the two contributions become equal also decreases with applied voltage, but saturation below the tungsten melting point occurs only for very low voltages; see Fig.\ \ref{Fig_I_B}(d). Here only up to 350 electrons are in the vacuum gap at each time, and thus small fluctuations of their number result in considerable noise.

In Fig.\ \ref{Fig_C_A} we show further the effects of space charge. In this figure the gap voltage is 50~V but the temperature is varied from 2100 to 3500~K. The left column of the figure shows the development, with temperature, of the electron density at the cathode using a normalized color map where darker colors imply higher density. For lower temperatures, the density reflects the underlying work function and the checkered pattern remains apparent, but, as the temperature is increased, the density becomes more uniform except that it is more pronounced at the emitter edges, as would be expected from previous studies of space-charge-limited current from a finite emitter \cite{Umstattd01}. Note that, for  $T = 2100$~K, the emission rate is low enough that statistical flutter partially obfuscates the checkerboard structure, which becomes more apparent at $T = 2500$~K. In Fig.\ \ref{Fig_Diagonal} we show the partial currents coming from the areas marked A, B, C, a, b, and c in Fig.\ \ref{Fig_Cathode}. The upper part of the figure shows the partial currents from low work function checks, and indicates that the current from check A, which is at the emitter edge, is higher than that from interior checks B and C, and that space-charge saturation occurs in the interior before it does so at the edge. The lower part of Fig.\ \ref{Fig_Diagonal} shows the partial currents from high work function checks. We note that, as for the low work function case, the partial currents from the interior checks are almost identical \red{and lower than that from the edge check; see Fig.\ \ref{Fig_Diagonal}(b)}.

%%%%%%%%%%%%%%%%%%%%%%%%%%%%%%%%%%%%%%%%%%%%%%%%%%%%%%%%%%%%%%%%%%%%%%%%%%%
\begin{figure}[t]
	\includegraphics[scale=0.67]{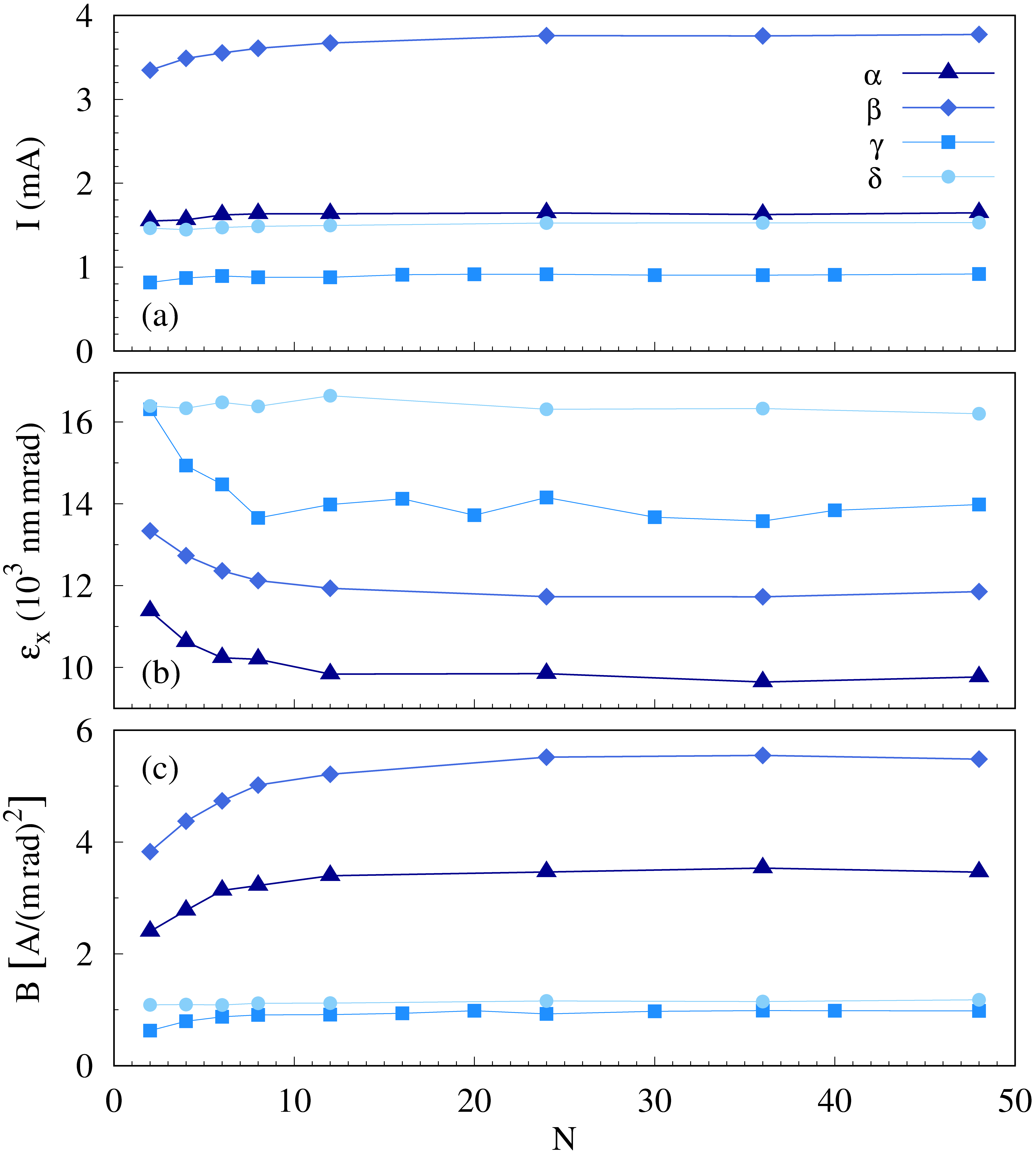}
	\caption{(a) Steady-state current, (b) the $x$ component of emittance, and (c) brightness  versus the number of checks for the four points indicated in Figs.\ \ref{Fig_I_B}(a) and\ \ref{Fig_I_B}(b). The values of applied bias voltage and temperature associated with the Greek letters are listed in Table I. The line description shown in panel (a) is valid for all panels.}
	\label{Fig_NxN}
\end{figure}
%%%%%%%%%%%%%%%%%%%%%%%%%%%%%%%%%%%%%%%%%%%%%%%%%%%%%%%%%%%%%%%%%%%%%%%%%%%

In the right column of Fig.\ \ref{Fig_C_A} we show the local density of electrons at the anode. Up to $T=2500$ K the beam envelope hardly changes, but the initial checkerboard structure of the 
beam is destroyed; see Figs.\ \ref{Fig_C_A}(b) and \ref{Fig_C_A}(d). The beamlets originating from  areas of lower work function spread and squeeze the beamlets formed by electrons released from  checks of higher work function. With increasing current density, the space-charge effect affects the beam profile in two ways.
On the one hand, it widens and rounds the beam envelope, and on the other 
hand, it pushes electrons towards the beam center.
Although, somewhat deformed, the original square shape of the emitting area can still be recognized at $T=2100$ and $2500$ K. At $T=3000$~K the beam cross section becomes more circular with a large fraction of electrons concentrated in its center; see Fig.\ \ref{Fig_C_A}(f).  At $T=3500$ K the beam has a circular cross section with the diameter approximately twice as large as the emitter edge. The density of electrons forms a sharp peak centered in the middle 
of the beam and monotonically decreases outward; see Fig.\ \ref{Fig_C_A}(h). Interestingly, at such high temperatures the largest number of electrons is emitted from the cathode boundaries [see Fig.\ \ref{Fig_C_A}(g)], but the space-charge forces totally redistribute them during the $0.3$ ps transit between the electrodes.

%=========================================================
\begin{center}
	\begin{table}[t]
		\begin{tabular}{K{2.5cm}||K{1.0cm}|K{1.0cm}|K{1.0cm}|K{1.0cm}} 
			Symbol     & $\alpha$ & $\beta$ & $\gamma$ & $\delta$  \\ \hline
			$V_0$ (V) & 100      & 100     & 50       & 50        \\ \hline
			$T$ (K)   & 2500     & 3000    & 2500     & 3000 
			\label{T_ABCD}                
		\end{tabular} 
		\caption{Applied bias voltages and temperatures for the line descriptions in Fig.\ \ref{Fig_NxN}.}  
	\end{table}		
\end{center}
%========================================================= 

To characterize the beam, we calculate its emittance and brightness. The former increases nearly linearly with temperature for the range that is under study, as can be seen in Figs\ \ref{Fig_E}. Once the number of electrons in the system is large enough to mask statistical fluctuation, the transverse emittance is seen to be symmetric. The brightness is a ratio between the emitted current and the emittance. Initially, all three parameters increase with temperature. When the space-charge effects become relevant, the increase in the current, and thus also in the brightness, slows down. Interestingly, the brightness attains a maximum value approximately at the peak value of the partial current from the low work function area, as can be seen in Figs.\ \ref{Fig_I_B}(e)-\ref{Fig_I_B}(h). This means that the optimal operational point for beam brightness is in the "knee" region of the Miram curve.

\subsection{\label{sec:grid} Grid effects}

%%%%%%%%%%%%%%%%%%%%%%%%%%%%%%%%%%%%%%%%%%%%%%%%%%%%%%%%%%%%%%%%%%%%%%%%%%%
\begin{figure}
	\includegraphics[scale=0.51]{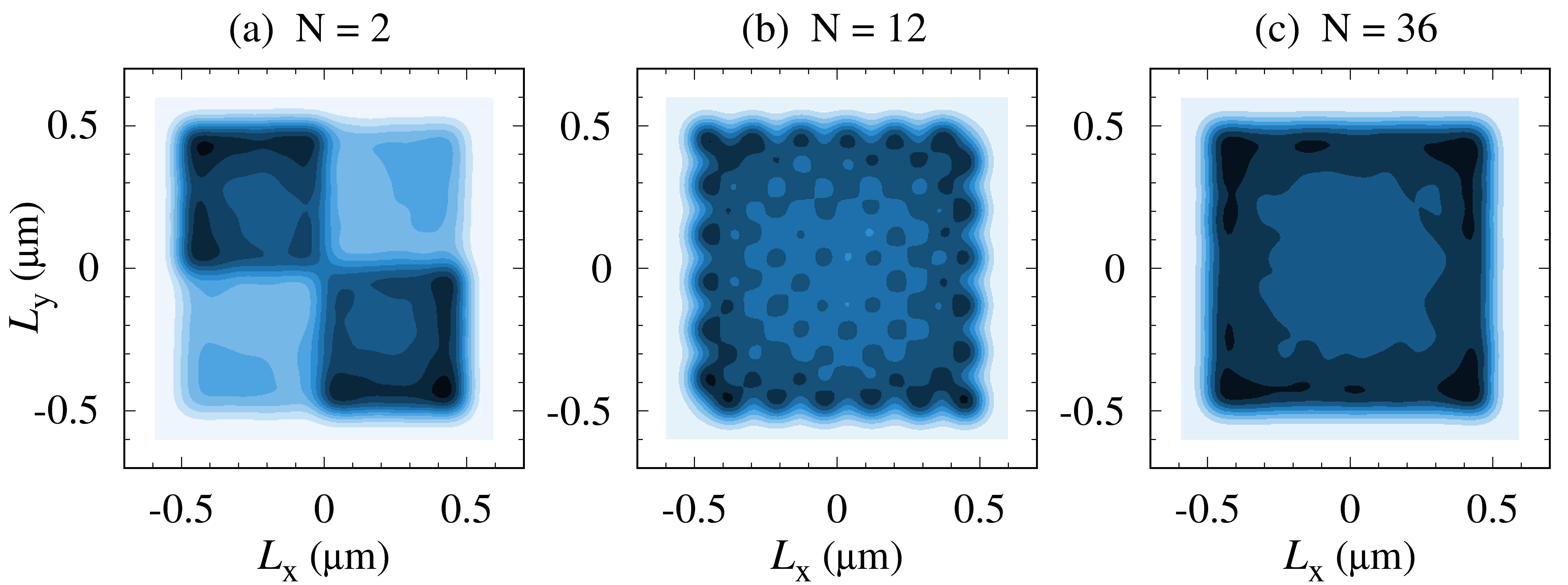}
	\caption{Local density of electrons at the cathode for the \red{applied voltage and temperate associated with the letter $\beta$ (Table I)} and for a different number of checks in each row and column: (a) $N=2$, (b) $N=12$, and (c) $N=36$.}
	\label{Fig_Dens_Cathode_N}
\end{figure}
%%%%%%%%%%%%%%%%%%%%%%%%%%%%%%%%%%%%%%%%%%%%%%%%%%%%%%%%%%%%%%%%%%%%%%%%%%%

Finally, we show how the fine-grain structure of the checkerboard affects the beam properties. We do this by varying the number of checks on the emitter while maintaining the edge length. We present results for the four different combinations of temperature and gap voltage listed in Table I, which are also shown in Figs.\ \ref{Fig_I_B}(a) and\ \ref{Fig_I_B}(b). In all studied cases the current and brightness increase with the number of checks [see Figs.\ \ref{Fig_NxN}(a) and\ \ref{Fig_NxN}(c)], while the emittance decreases [see Fig.\ \ref{Fig_NxN}(b)]. When the \red{emitting} surface is divided into four checks, the electron distribution is structurally similar to that which would be found in four independent emitters of uniform work function, except that we see diminished emission from the corners of the low work function checks that abut each other at the center of the emitter. This is due to mutual space-charge effects between those two checks. Mutual space-charge effects between adjacent emitters are known to vary rapidly with decreasing distance between them \cite{Haraldsson20,Ilkov15a}, and as the number of checks is increased, the effect is to decrease the spacing between \red{them}, causing stronger space-charge coupling until the emitting area may be considered to be practically uniform and further increase in the number of checks has no effect. Indeed, this can be seen in Fig.\ \ref{Fig_Dens_Cathode_N}(c) where the electron density at the cathode shows the same type of symmetric structure as would be the case for a cathode of uniform work function. It should be noted that the current, emittance, and brightness from this "apparently uniform" emitter with a large number of checks is neither equivalent to that of a uniformly low or high work function, nor that of their average value. 
In fact, the current that can be drawn from a fine-grained emitter will always be less than that drawn from a uniform emitter of low work function.

\section{\label{sec:conclusions} Conclusions}

In this paper we examine current and beam characteristics for field-assisted thermionic emission from an emitter of finite dimension and inhomogeneous work function in a planar diode. Our results show that, with increasing temperature, the current initially comes from areas of low work function, even to such an extent that the Child-Langmuir current density is locally exceeded, but eventually it becomes evenly distributed across the  \red{center of the} emitting area of the cathode, irrespective of local work function. This is in accordance with results obtained by Chernin \textit{et al.} \cite{Chernin20}. However, we also demonstrate that the soft "knee" in transition from the source-limited to space-charge-limited parts of the Miram curve can be obtained from a cathode of homogeneous work function. This is different from the results of Chernin \textit{et al.} \cite{Chernin20}, and is most likely due to the fact that in our case the emitter area is of finite size as compared to their model in which periodic boundaries are used so that the emitting area is effectively infinite. Thus, edge effects and transverse beam expansion are likely to influence results. 
\red{In Longo \cite{Longo03} the smooth transition from thermionic to space-charge-limited emission is related to a shape factor, $\alpha$, which is dependent on the disorder of the emitting surface of the cathode. With the data described in this paper, it is not possible to discern the edge effects, due to limited emitter size, completely from the effects of varying work function.}

We show how emittance and brightness of the beam from the cathode are dependent on the \red{temperature}, and that there is an optimal temperature value to maximize the brightness. This coincides with the temperature at which the partial current from low work function regions is maximized in the "knee" of the Miram curve. We also show how the current, brightness, and emittance are affected by how fine grained the checkerboard pattern of the work function is. Brightness and current increase to an asymptotic value with increasing fineness of the cathode patterning. Nonetheless, the source-limited current from a cathode with alternating low and high work function checks is always lower than that from a cathode with uniform low work function at the same temperature, irrespective of how fine grained it is.

\begin{acknowledgments}
This work is supported by the Air Force Office of Scientific Research under Grant No. FA9550-18-1-7011, and by the Icelandic Research Fund Grant No. 174127-051. 
Any opinions, findings, and conclusions or recommendations expressed in this material are those of the authors and do not necessarily reflect the views of the United States Air Force.
\end{acknowledgments}

%\bibliography{vacuum.bib}

\begin{thebibliography}{17}%
	\makeatletter
	\providecommand \@ifxundefined [1]{%
		\@ifx{#1\undefined}
	}%
	\providecommand \@ifnum [1]{%
		\ifnum #1\expandafter \@firstoftwo
		\else \expandafter \@secondoftwo
		\fi
	}%
	\providecommand \@ifx [1]{%
		\ifx #1\expandafter \@firstoftwo
		\else \expandafter \@secondoftwo
		\fi
	}%
	\providecommand \natexlab [1]{#1}%
	\providecommand \enquote  [1]{``#1''}%
	\providecommand \bibnamefont  [1]{#1}%
	\providecommand \bibfnamefont [1]{#1}%
	\providecommand \citenamefont [1]{#1}%
	\providecommand \href@noop [0]{\@secondoftwo}%
	\providecommand \href [0]{\begingroup \@sanitize@url \@href}%
	\providecommand \@href[1]{\@@startlink{#1}\@@href}%
	\providecommand \@@href[1]{\endgroup#1\@@endlink}%
	\providecommand \@sanitize@url [0]{\catcode `\\12\catcode `\$12\catcode
		`\&12\catcode `\#12\catcode `\^12\catcode `\_12\catcode `\%12\relax}%
	\providecommand \@@startlink[1]{}%
	\providecommand \@@endlink[0]{}%
	\providecommand \url  [0]{\begingroup\@sanitize@url \@url }%
	\providecommand \@url [1]{\endgroup\@href {#1}{\urlprefix }}%
	\providecommand \urlprefix  [0]{URL }%
	\providecommand \Eprint [0]{\href }%
	\providecommand \doibase [0]{http://dx.doi.org/}%
	\providecommand \selectlanguage [0]{\@gobble}%
	\providecommand \bibinfo  [0]{\@secondoftwo}%
	\providecommand \bibfield  [0]{\@secondoftwo}%
	\providecommand \translation [1]{[#1]}%
	\providecommand \BibitemOpen [0]{}%
	\providecommand \bibitemStop [0]{}%
	\providecommand \bibitemNoStop [0]{.\EOS\space}%
	\providecommand \EOS [0]{\spacefactor3000\relax}%
	\providecommand \BibitemShut  [1]{\csname bibitem#1\endcsname}%
	\let\auto@bib@innerbib\@empty
%	</preamble>
	\bibitem [{\citenamefont {Dowell}\ \emph {et~al.}(2010)\citenamefont {Dowell},
		\citenamefont {Bazarov}, \citenamefont {Dunham}, \citenamefont {Harkay},
		\citenamefont {Hernandez-Garcia}, \citenamefont {Legg}, \citenamefont
		{Padmore}, \citenamefont {Rao}, \citenamefont {Smedley},\ and\ \citenamefont
		{Wan}}]{Dowell10}%
	\BibitemOpen
	\bibfield  {author} {\bibinfo {author} {\bibfnamefont {D.}~\bibnamefont
			{Dowell}}, \bibinfo {author} {\bibfnamefont {I.}~\bibnamefont {Bazarov}},
		\bibinfo {author} {\bibfnamefont {B.}~\bibnamefont {Dunham}}, \bibinfo
		{author} {\bibfnamefont {K.}~\bibnamefont {Harkay}}, \bibinfo {author}
		{\bibfnamefont {C.}~\bibnamefont {Hernandez-Garcia}}, \bibinfo {author}
		{\bibfnamefont {R.}~\bibnamefont {Legg}}, \bibinfo {author} {\bibfnamefont
			{H.}~\bibnamefont {Padmore}}, \bibinfo {author} {\bibfnamefont
			{T.}~\bibnamefont {Rao}}, \bibinfo {author} {\bibfnamefont {J.}~\bibnamefont
			{Smedley}}, \ and\ \bibinfo {author} {\bibfnamefont {W.}~\bibnamefont
			{Wan}},\ }
%
%
%	\bibitem [{\citenamefont {Dowell}(2010)}]{Dowell10}%
%\BibitemOpen
%\bibfield  {author} {\bibinfo {author} {\bibfnamefont {D.}~\bibnamefont
%		{Dowell}},\ \ \emph {et~al.}, }
%
%
		\href {\doibase https://doi.org/10.1016/j.nima.2010.03.104}
	{\bibfield  {journal} {\bibinfo  {journal} {Nucl. Instrum. Meth.
				Phys. Res., Sect. A}\ }\textbf {\bibinfo {volume} {622}},\ \bibinfo {pages}
		{685 } (\bibinfo {year} {2010})}\BibitemShut {NoStop}%
%
%		
	\bibitem [{\citenamefont {Zhang}\ \emph {et~al.}(2017)\citenamefont {Zhang},
		\citenamefont {Valfells}, \citenamefont {Ang}, \citenamefont {Luginsland},\
		and\ \citenamefont {Lau}}]{Zhang17a}%
	\BibitemOpen
	\bibfield  {author} {\bibinfo {author} {\bibfnamefont {P.}~\bibnamefont
			{Zhang}}, \bibinfo {author} {\bibfnamefont {A.}~\bibnamefont {Valfells}},
		\bibinfo {author} {\bibfnamefont {L.~K.}\ \bibnamefont {Ang}}, \bibinfo
		{author} {\bibfnamefont {J.~W.}\ \bibnamefont {Luginsland}}, \ and\ \bibinfo
		{author} {\bibfnamefont {Y.~Y.}\ \bibnamefont {Lau}},\ }\href {\doibase
		10.1063/1.4978231} {\bibfield  {journal} {\bibinfo  {journal} {Appl.
				Phys. Rev.}\ }\textbf {\bibinfo {volume} {4}},\ \bibinfo {pages}
		{011304} (\bibinfo {year} {2017})}.\ \Eprint
	{http://arxiv.org/abs/https://doi.org/10.1063/1.4978231}
%	{https://doi.org/10.1063/1.4978231} 
\BibitemShut %
	\bibitem [{\citenamefont {Richardson}(1916)}]{Richardson16}%
	\BibitemOpen
	\bibfield  {author} {\bibinfo {author} {\bibfnamefont {O.~W.}\ \bibnamefont
			{Richardson}},\ }\href@noop {} {\emph {\bibinfo {title} {The Emission of
				Electricity from Hot Bodies}}}\ (\bibinfo  {publisher} {London, U.K.:
		Longmans, Green, and Company},\ \bibinfo {year} {1916})\BibitemShut {NoStop}%
	\bibitem [{\citenamefont {Dushman}(1923)}]{Dushman23}%
	\BibitemOpen
	\bibfield  {author} {\bibinfo {author} {\bibfnamefont {S.}~\bibnamefont
			{Dushman}},\ }\href {\doibase 10.1103/PhysRev.21.623} {\bibfield  {journal}
		{\bibinfo  {journal} {Phys. Rev.}\ }\textbf {\bibinfo {volume} {21}},\
		\bibinfo {pages} {623} (\bibinfo {year} {1923})}\BibitemShut {NoStop}%
	\bibitem [{\citenamefont {{Chernin}}\ \emph {et~al.}(2020)\citenamefont
		{{Chernin}}, \citenamefont {{Lau}}, \citenamefont {{Petillo}}, \citenamefont
		{{Ovtchinnikov}}, \citenamefont {{Chen}}, \citenamefont {{Jassem}},
		\citenamefont {{Jacobs}}, \citenamefont {{Morgan}},\ and\ \citenamefont
		{{Booske}}}]{Chernin20}%
	\BibitemOpen
	\bibfield  {author} {\bibinfo {author} {\bibfnamefont {D.}~\bibnamefont
			{{Chernin}}}, \bibinfo {author} {\bibfnamefont {Y.~Y.}\ \bibnamefont
			{{Lau}}}, \bibinfo {author} {\bibfnamefont {J.~J.}\ \bibnamefont
			{{Petillo}}}, \bibinfo {author} {\bibfnamefont {S.}~\bibnamefont
			{{Ovtchinnikov}}}, \bibinfo {author} {\bibfnamefont {D.}~\bibnamefont
			{{Chen}}}, \bibinfo {author} {\bibfnamefont {A.}~\bibnamefont {{Jassem}}},
		\bibinfo {author} {\bibfnamefont {R.}~\bibnamefont {{Jacobs}}}, \bibinfo
		{author} {\bibfnamefont {D.}~\bibnamefont {{Morgan}}}, \ and\ \bibinfo
		{author} {\bibfnamefont {J.~H.}\ \bibnamefont {{Booske}}},\ }\href@noop {}
	{\bibfield  {journal} {\bibinfo  {journal} {IEEE Trans. Plasma
				Sci.}\ }\textbf {\bibinfo {volume} {48}},\ \bibinfo {pages} {146}
		(\bibinfo {year} {2020})}\BibitemShut {NoStop}%
	\bibitem [{\citenamefont {Jensen}(2018)}]{Kevin}%
	\BibitemOpen
	\bibfield  {author} {\bibinfo {author} {\bibfnamefont {K.~L.}\ \bibnamefont
			{Jensen}},\ }\href@noop {} {\emph {\bibinfo {title} {Introduction to the
				Physics of Electron Emission}}}\ (\bibinfo  {publisher} {John Wiley and
		Sons},\ \bibinfo {address} {Hoboken, New Jersey},\ \bibinfo {year}
	{2018})\BibitemShut {NoStop}%
	\bibitem [{\citenamefont {{Ramo}}(1939)}]{Ramo39}%
	\BibitemOpen
	\bibfield  {author} {\bibinfo {author} {\bibfnamefont {S.}~\bibnamefont
			{{Ramo}}},\ }\href {\doibase 10.1109/JRPROC.1939.228757} {\bibfield
		{journal} {\bibinfo  {journal} {Proc. IRE}\ }\textbf {\bibinfo
			{volume} {27}},\ \bibinfo {pages} {584} (\bibinfo {year} {1939})}\BibitemShut
	{NoStop}%
	\bibitem [{\citenamefont {Reiser}(1994)}]{Reiser94}%
	\BibitemOpen
	\bibfield  {author} {\bibinfo {author} {\bibfnamefont {M.}~\bibnamefont
			{Reiser}},\ }\href {\doibase 10.1002/9783527622047} {\emph {\bibinfo {title}
			{Theory and design of charged particle beams}}}\ (\bibinfo  {publisher}
	{Wiley Online Library},\ \bibinfo {year} {1994})\BibitemShut {NoStop}%
	\bibitem [{\citenamefont {Buon}(1992)}]{buon1992beam}%
	\BibitemOpen
	\bibfield  {author} {\bibinfo {author} {\bibfnamefont {J.}~\bibnamefont
			{Buon}},\ }\href@noop {} {\emph {\bibinfo {title} {Beam phase space and
				emittance (LAL-RT--92-03)}}},\ \bibinfo {type} {Tech. Rep.}\ (\bibinfo
	{institution} {Paris-11 Univ.},\ \bibinfo {year} {1992})\BibitemShut
	{NoStop}%
	\bibitem [{\citenamefont {Torfason}\ \emph {et~al.}(2015)\citenamefont
		{Torfason}, \citenamefont {Valfells},\ and\ \citenamefont
		{Manolescu}}]{Torfason15}%
	\BibitemOpen
	\bibfield  {author} {\bibinfo {author} {\bibfnamefont {K.}~\bibnamefont
			{Torfason}}, \bibinfo {author} {\bibfnamefont {A.}~\bibnamefont {Valfells}},
		\ and\ \bibinfo {author} {\bibfnamefont {A.}~\bibnamefont {Manolescu}},\
	}\href {\doibase 10.1063/1.4914855} {\bibfield  {journal} {\bibinfo
			{journal} {Phys. Plasmas}\ }\textbf {\bibinfo {volume} {22}},\ \bibinfo
		{pages} {033109} (\bibinfo {year} {2015})}.\ \Eprint
	{http://arxiv.org/abs/https://doi.org/10.1063/1.4914855}
%	{https://doi.org/10.1063/1.4914855} 
\BibitemShut %
	\bibitem [{\citenamefont {Torfason}\ \emph {et~al.}(2020)\citenamefont
		{Torfason}, \citenamefont {Sitek}, \citenamefont {Manolescu},\ and\
		\citenamefont {Valfells}}]{Torfason20}%
	\BibitemOpen
	\bibfield  {author} {\bibinfo {author} {\bibfnamefont {K.}~\bibnamefont
			{Torfason}}, \bibinfo {author} {\bibfnamefont {A.}~\bibnamefont {Sitek}},
		\bibinfo {author} {\bibfnamefont {A.}~\bibnamefont {Manolescu}}, \ and\
		\bibinfo {author} {\bibfnamefont {A.}~\bibnamefont {Valfells}},\ }\href@noop
	{} {\bibfield  {journal} {\bibinfo  {journal} {arXiv:2011.13731}\ } (\bibinfo
		{year} {2020})}\BibitemShut {NoStop}%
	\bibitem [{\citenamefont {Lau}(2001)}]{Lau01}%
	\BibitemOpen
	\bibfield  {author} {\bibinfo {author} {\bibfnamefont {Y.~Y.}\ \bibnamefont
			{Lau}},\ }\href {\doibase 10.1103/PhysRevLett.87.278301} {\bibfield
		{journal} {\bibinfo  {journal} {Phys. Rev. Lett.}\ }\textbf {\bibinfo
			{volume} {87}},\ \bibinfo {pages} {278301} (\bibinfo {year}
		{2001})}\BibitemShut {NoStop}%
	\bibitem [{\citenamefont {Luginsland}\ \emph {et~al.}(1996)\citenamefont
		{Luginsland}, \citenamefont {Lau},\ and\ \citenamefont
		{Gilgenbach}}]{Luginsland96}%
	\BibitemOpen
	\bibfield  {author} {\bibinfo {author} {\bibfnamefont {J.~W.}\ \bibnamefont
			{Luginsland}}, \bibinfo {author} {\bibfnamefont {Y.~Y.}\ \bibnamefont {Lau}},
		\ and\ \bibinfo {author} {\bibfnamefont {R.~M.}\ \bibnamefont {Gilgenbach}},\
	}\href {\doibase 10.1103/PhysRevLett.77.4668} {\bibfield  {journal} {\bibinfo
			{journal} {Phys. Rev. Lett.}\ }\textbf {\bibinfo {volume} {77}},\ \bibinfo
		{pages} {4668} (\bibinfo {year} {1996})}\BibitemShut {NoStop}%
	\bibitem [{\citenamefont {Umstattd}\ and\ \citenamefont
		{Luginsland}(2001)}]{Umstattd01}%
	\BibitemOpen
	\bibfield  {author} {\bibinfo {author} {\bibfnamefont {R.~J.}\ \bibnamefont
			{Umstattd}}\ and\ \bibinfo {author} {\bibfnamefont {J.~W.}\ \bibnamefont
			{Luginsland}},\ }\href {\doibase 10.1103/PhysRevLett.87.145002} {\bibfield
		{journal} {\bibinfo  {journal} {Phys. Rev. Lett.}\ }\textbf {\bibinfo
			{volume} {87}},\ \bibinfo {pages} {145002} (\bibinfo {year}
		{2001})}\BibitemShut {NoStop}%
	\bibitem [{\citenamefont {Haraldsson}\ \emph {et~al.}(2020)\citenamefont
		{Haraldsson}, \citenamefont {Torfason}, \citenamefont {Manolescu},\ and\
		\citenamefont {Valfells}}]{Haraldsson20}%
	\BibitemOpen
	\bibfield  {author} {\bibinfo {author} {\bibfnamefont {H.~V.}\ \bibnamefont
			{Haraldsson}}, \bibinfo {author} {\bibfnamefont {K.}~\bibnamefont
			{Torfason}}, \bibinfo {author} {\bibfnamefont {A.}~\bibnamefont {Manolescu}},
		\ and\ \bibinfo {author} {\bibfnamefont {A.}~\bibnamefont {Valfells}},\
	}\href@noop {} {\bibfield  {journal} {\bibinfo  {journal} {IEEE Trans.
				Plasma Sci.}\ }\textbf {\bibinfo {volume} {48}},\ \bibinfo {pages}
		{1967} (\bibinfo {year} {2020})}\BibitemShut {NoStop}%
	\bibitem [{\citenamefont {Ilkov}\ \emph {et~al.}(2015)\citenamefont {Ilkov},
		\citenamefont {Torfason}, \citenamefont {Manolescu},\ and\ \citenamefont
		{Valfells}}]{Ilkov15a}%
	\BibitemOpen
	\bibfield  {author} {\bibinfo {author} {\bibfnamefont {M.}~\bibnamefont
			{Ilkov}}, \bibinfo {author} {\bibfnamefont {K.}~\bibnamefont {Torfason}},
		\bibinfo {author} {\bibfnamefont {A.}~\bibnamefont {Manolescu}}, \ and\
		\bibinfo {author} {\bibfnamefont {A.}~\bibnamefont {Valfells}},\ }\href@noop
	{} {\bibfield  {journal} {\bibinfo  {journal} {IEEE Trans. Electron
				Devices}\ }\textbf {\bibinfo {volume} {62}},\ \bibinfo {pages} {200}
		(\bibinfo {year} {2015})}\BibitemShut {NoStop}%
	\bibitem [{\citenamefont {Longo}(2003)}]{Longo03}%
	\BibitemOpen
	\bibfield  {author} {\bibinfo {author} {\bibfnamefont {R.~T.}\ \bibnamefont
			{Longo}},\ }\href {\doibase 10.1063/1.1621728} {\bibfield  {journal}
		{\bibinfo  {journal} {J. Appl. Phys.}\ }\textbf {\bibinfo
			{volume} {94}},\ \bibinfo {pages} {6966} (\bibinfo {year} {2003})}.\ \Eprint
	{http://arxiv.org/abs/https://doi.org/10.1063/1.1621728}
%	{https://doi.org/10.1063/1.1621728}
 \BibitemShut %
\end{thebibliography}


\end{document}